# Aluminum Josephson junction microstructure and electrical properties modification with thermal annealing

**Nikita D. Korshakov[1,2], Dmitry O. Moskalev[1,2], Anastasiya A. Soloviova[1,2], Daria A. Moskaleva[1,2], Evgeniy S. Lotkov[1,2], Artyom R. Ibragimov[1], Margarita V. Androschuk[1], Ilya A. Ryzhikov [1,3], Yuri V. Panfilov [1] and Ilya A. Rodionov[1,2,*]**

[1] FMN Laboratory, Bauman Moscow State Technical University, Moscow, 105005, Russia
[2] Dukhov Automatics Research Institute (VNIIA), Moscow, 127055, Russia
[3] Institute of Theoretical and Applied Electrodynamics, RAS, Moscow, 125412, Russia
[*]Electronic mail: irodionov@bmstu.ru

**ABSTRACT**

Superconducting qubits based on Al/AlO$_x$/Al Josephson junction are one of the most promising candidates for the physical implementation of universal quantum computers. Due to scalability and compatibility with the state-of-the-art nanoelectronic processes one can fabricate hundreds of qubits on a single silicon chip. However, decoherence in these systems caused by two-level-systems in amorphous dielectrics, including a tunneling barrier AlO$_x$, is one of the major problems. We report on a Josephson junction thermal annealing process development to crystallize an amorphous barrier oxide (AlO$_x$). The dependences of the thermal annealing parameters on the room temperature resistance are obtained. The developed method allows not only to increase the Josephson junction resistance by 175%, but also to decrease by 60% with precisions of 10% in R$_n$. Finally, theoretical assumptions about the structure modification in tunnel barrier are proposed. The suggested thermal annealing approach can be used to form a stable and reproducible tunneling barriers and scalable frequency trimming for a widely used fixed-frequency transmon qubits.

**Introduction**

Superconducting quantum circuits is a promising solution for logic quantum gates realization and quantum simulators [1-7]. However, one of the most serious obstacles to the realization of a quantum computer is the frequency collisions when scaling qubit circuits and the decoherence of these systems [8].

Superconducting qubit is strongly coupled to its environment, making it more susceptible to uncontrolled decoherence sources [9, 10]. For example, coupling to the outside world occurs through the electrical leads and to the device-level surroundings through direct electromagnetic interactions. Decoherence can be suppressed by means of filtering and shielding techniques [11]. At the device level, progress has been made by optimizing the circuit design operating the qubit at an optimal bias point or using special measurement techniques [12-15]. However, such approaches may not be effective when decoherence sources originate within material components forming the qubit.

The principal challenge for scaling fixed-frequency architectures is errors arising from lattice frequency collisions. Typical fabrication tolerances for transmon qubits frequencies range from 1 to 2%, with uncertainties dominated by the 2 to 4% variation in tunnel junction resistance (R$_n$) [16, 17].

The solution to the problem can be thermal annealing on superconducting quantum circuits. Thermal annealing methods to adjust and stabilize post-fabrication R$_n$ (and correspondingly, transmon frequencies f$_{01}$) have been explored previously [18-21]. More recently, the LASIQ (Laser Annealing of Stochastically Impaired Qubits) technique was introduced to increase collision-free yield of transmon lattices by selectively trimming (i.e., tuning) individual qubit frequencies via laser thermal annealing [16]. The LASIQ process sets R$_n$ with high precision, and f$_{01}$ could be predicted from R$_n$ according to a power-law relationship resulting from the Ambegaokar-Baratoff relations and transmon theory [22, 23]. Changes in electrical characteristics occur due to the modification of the Josephson junction barrier structure (crystallization of amorphous AlO$_x$). This makes it possible to improve the coherence of superconducting circuits by reducing the density of two-level systems (TLS) [8].

In this work, we demonstrate a method of thermal annealing on Josephson junctions of superconducting qubits. This method allows frequency detuning of qubits. The obtained results open up new possibilities for controlling the frequencies of superconducting qubit. Finally, the theoretical hypotheses put forward make it possible to understand the internal mechanisms of the tunnel barrier under thermal exposure, which affect the electrical characteristics.

**Experimental details**

For this study, we used high-resistivity silicon substrates (10,000 Ω cm). Prior to the base layer deposition, the substrate is cleaned in a Piranha solution at 80°C, followed by dipping in hydrofluoric bath [24]. 100 nm thick Al base layer is deposited using ultra high vacuum e-beam evaporation system. Pads were defined using a direct-laser lithography and dry-etched in $BCl_3/Cl_2$ inductively coupled plasma. The Josephson junctions (JJ), described in this work, were fabricated using Niemeyer-Dolan technique [25, 26]. This method has several advantages over the Manhattan junctions [27, 28]. Smaller evaporation angles result in improved electrode surface roughness. Also, bridge technology is more suitable for fabrication of large arrays of junction, e.g., for parametric amplifiers [29]. The substrate is spin coated with resist bilayer composed of EL9 copolymer and chemically amplified resist CSAR 62. Layouts were generated and exposed with 50 keV e-beam lithography system. $Al/AlO_x/Al$ junctions are shadow evaporated in ultra-high vacuum deposition system. Resist lift-off was performed in N-methyl-2-pyrrolidone. Finally, we patterned and evaporated aluminum bandages using the same process as for junctions with an in-situ Ar ion milling [30].

The chip topology contains an JJ array with areas of 0.01, 0.025 and 0.1 $\mu m^2$. The topology additionally includes junctions with different ratios of the top and bottom electrodes linear dimensions. Linear dimensions and contribution of area sidewall are presented in Table 1.

| Top electrode, nm | Bottom electrode, nm | Contribution[1], % |
|---|---|---|
| 150 | 170 | 39 |
| 160 | 160 | 41 |
| 190 | 130 | 46 |
| 230 | 110 | 50 |
| 150 | 670 | 14 |
| 210 | 480 | 19 |
| 260 | 390 | 22 |
| 320 | 320 | 26 |

[1] The contribution is calculated $\frac{2 \times Top \times h}{Top \times Bottom + 2 \times Top \times h} \times 100\%$, where h – thickness bottom electrode

**Table 1.** The top and bottom electrodes linear dimensions in topology.

The quality and uniformity of the deposited electrodes was examined using a scanning electron microscopy. To investigate the microstructure of Josephson junctions, lamellae are prepared and then examined using the facility of TEM, which is equipped with a x-ray Energy Dispersive Spectroscopy (EDS). The Josephson junctions room temperature resistance were individually measured with automated probe station. The stylus profiler was used to measure surface roughness of bottom JJ electrode. Thermal annealing was carried out in a rapid thermal process powerful multi-zone infrared lamp furnace. The substrate temperature was controlled using thermocouples. Annealing was carried out in an argon atmosphere with varying temperature and holding time.

**Experimental results and discussion**

All samples were subjected to thermal annealing with the same heating time (10 minutes) and cooling time (60 minutes). Processes were carried out with varying temperatures (200, 300, 400, and 500°C) and holding time at temperature (w/o holding time, 10 and 60 minutes). Typical process for 400°C and 10 minutes holding time are shown in Figure 1a. Dependences $R_n$ on the annealing temperature and holding time are shown in Figure 1e-g. We additionally determined the difference between room resistances before and after thermal annealing (Δ).

For annealing at 200°C and different holding times (Fig. 1e-g), a decrease in $R_n$ by Δ=12-23% was observed depending on the junction area. For annealing at 300 and 400°C, the same trend remains as for 200°C. At 300°C with 60 minutes holding a decrease in $R_n$ by Δ=30% for all junction sizes. However, at 400°C temperature and 60 min holding time there is an increase in the $R_n$ for all Josephson junction sizes from 12% to 175% (Fig. 1d). For thermal annealing at a temperature of 500°C the thin-film coating of the Josephson junction was destroyed and the spread of resistance increased up to 40% over the chip. This effect indicates the impossibility of using temperatures above 400°C for junction thermal annealing.

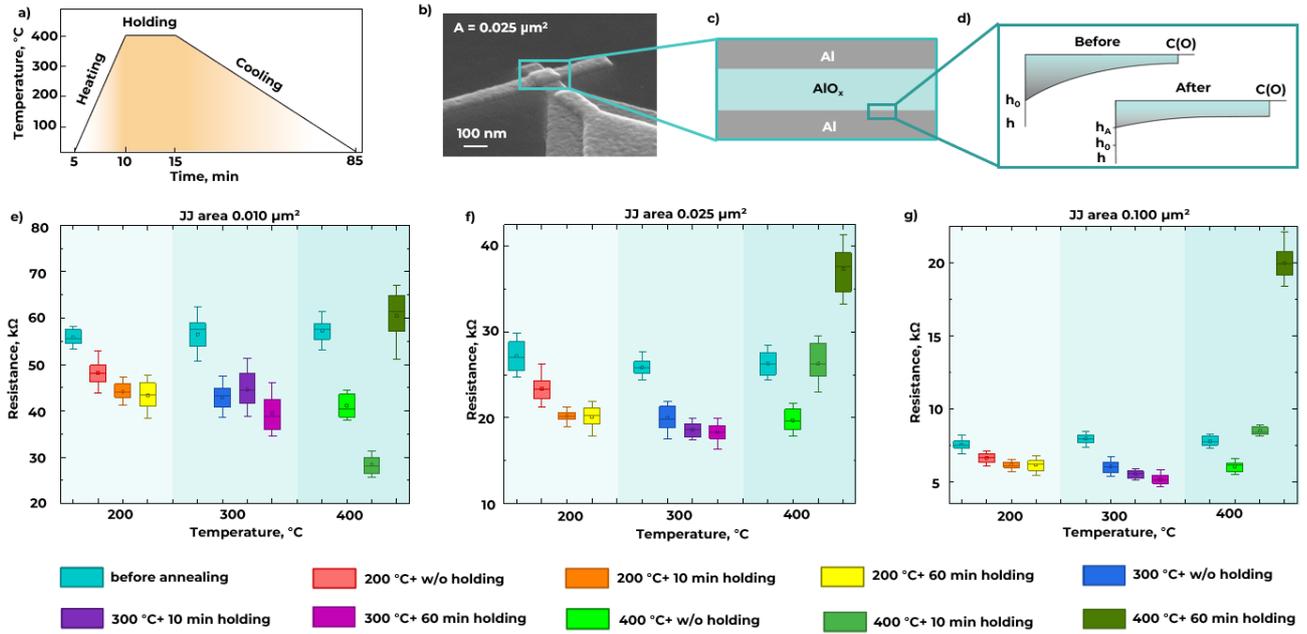

**Figure 1.** Typical process of thermal annealing with 10 minutes heating time, 10 minutes holding time and 60 minutes cooling time for 400°C (a); SEM image of fabricated junction used in this work (b); diagram of a single JJ (c); oxygen concentration gradient in the tunnel barrier at the Al/AlO$_x$ interface before and after thermal annealing (d); dependence of the change in the resistance value on the thermal annealing mode for JJ with area 0.010 μm$^2$ (e); dependence of the change in the resistance value on the thermal annealing mode for JJ with area 0.023 μm$^2$ (d); dependence of the change in the resistance value on the thermal annealing mode for JJ with area 0.100 μm$^2$ (g).

As a result of Josephson junctions thermal annealing, we obtained both an increase and a decrease in the room temperature resistance. The increase in resistance is a typical result and has been obtained in many works [31-33]. The decrease is an unexpected observation and can be theoretically described by the diffusion of oxygen atoms in the tunnel barrier. Figure 2a, b shows the tunneling barrier of the Josephson junction. As a result of the JJ aluminum electrode thermal oxidation, an oxygen concentration gradient is observed in the tunnel barrier (Fig. 2c) [35]. Due to thermal action and electric fields in the barrier of the Josephson junction, the oxygen concentration gradient decreases (Fig. 1d). This reduces the effective thickness of the tunnel barrier from $h_0$ to $h_A$, which entails a decrease resistance. Increase in resistance can be obtained by crystallizing the amorphous structure of aluminum oxide [20, 31-33, 35]. This result was noticed only in one mode of thermal annealing - 400°C and 60 min holding for all JJ sizes. This is explained in the following way. Before reaching a certain temperature and holding time, the system does not have enough energy for the transition from an amorphous state to a crystalline one. This threshold is reached at 400°C and holding times in the range of 10-60 minutes.

With an increase in the contribution of sidewall (Table 1), differences in the change in resistance are observed (Fig. 2a). More specifically, for sizes 150x170 nm$^2$ (the sidewall contribution is 39%) and 230x110 nm$^2$ (the sidewall contribution is 50%) with annealing mode 400C and holding time of 60 minutes, room resistance increased by 38% and 15%, respectively. To understand the physics of the process, lamellae cut from sites of Josephson junctions were investigated using TEM (Fig. 2b). To carry out chemical analysis, the technique of high-resolution EDS is applied to mapping the upper layer of tunnel barrier. TEM micrograph coupled with elemental maps of O and a cross-sectional profile of oxygen atomic percentage confirms the presence of an oxygen concentration gradient in the tunnel barrier. Instead of an abrupt change of oxygen content at the AlO$_x$/Al interface, interfacial layers with a thickness about 0.5 nm with oxygen deficiency are observed. A similar study carried out for sidewall of the Josephson junction of markedly high content of oxygen in this region (Fig. 2e). Sidewall with an obviously thicker oxide layer does not initially involve in tunneling process, as can be seen from the initial room resistance for all Josephson junction sizes (Fig. 2a). However, annealing at 400°C and 10 minutes of exposure activates the lateral diffusion mechanism. Oxygen from the more saturated sidewall areas begins to move to deficient areas, leveling the oxygen concentration inside the tunnel barrier. This increases the effective area of the Josephson junction involved in the tunneling process. As the contribution of the side walls to the Josephson junction area increases from 39% to 50%, the resulting area changes from 0.033 to 0.037 μm$^2$, respectively. This difference in the final area of the Josephson junction determines the difference in room resistance by 2.5 times for shapes 150x170 nm$^2$ and 230x110 nm$^2$ and the annealing mode at 400°C and 60 min holding.

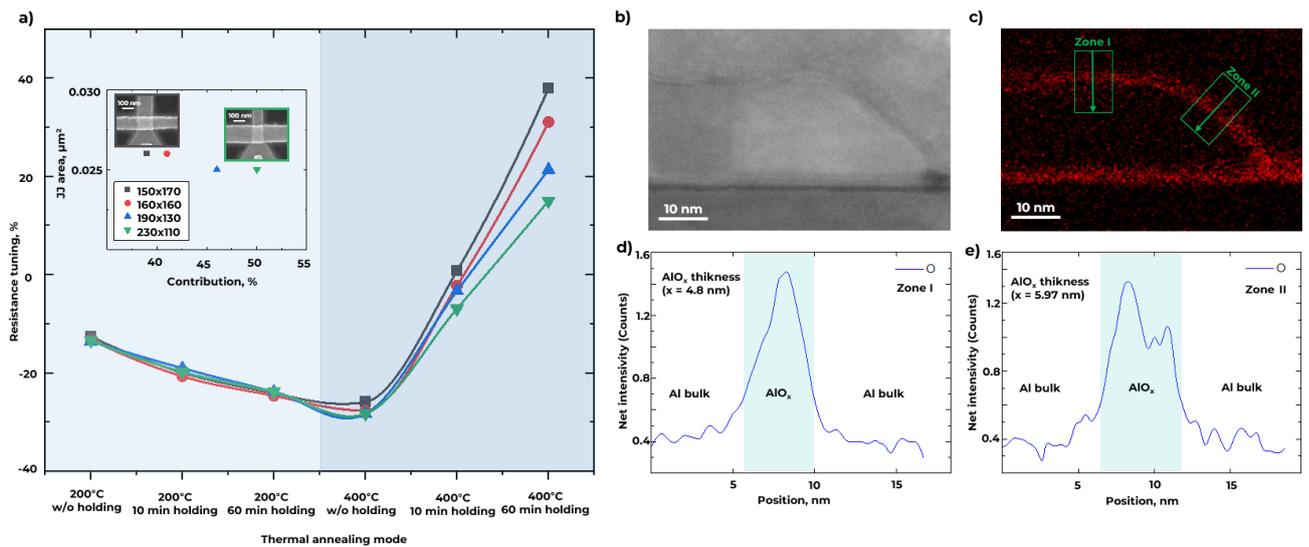

**Figure 2.** Dependence of the change in the resistance on the thermal annealing mode for JJ with different contribution of sidewall (junction area – 0.025 μm$^2$) (a); TEM image of the AlO$_x$ layer (b); quantitative map of oxygen (atomic percentage) (c); cross-sectional line profile of oxygen ratio from the marked area (zone I – center) (d); cross-sectional line profile of oxygen ratio from the marked area (zone II – sidewall) (e).

For the annealing mode of 200°C w/o holding and 400°C with 60 min holding time, the change in the root mean square roughness of the bottom electrode was estimated. Both mode lead to a change in the morphology of thin film coatings. For annealing at 200°C w/o holding did not give significant changes in roughness. Whereas annealing at 400°C with 60 min holding time led to an increase in the surface roughness of the bottom electrode by 4.4 A. Changing the bottom electrode morphology affects the effective area of the Josephson junction. Previous study shown that only 10% of AlO$_x$ tunnel barrier area of Josephson junction is actively participate in a tunneling process [36]. AlO$_x$ thickness variations are dominantly caused by grain boundary grooving in a bottom polycrystalline Al electrode. The tunnelling probability of charge carriers across the barrier is an exponential function of the barrier thickness. It has been shown that a 0.2 nm decrease of barrier thickness could result in one order of magnitude change in tunnelling current [37]. Thus, the thinnest region in the barrier may act as an active region or «hot spot» for tunnelling. Increasing the RMS surface roughness of the electrode may decrease the percentage of area involved in tunneling. Thus, decreasing the effective area increases $R_n$.

Changing the resistance in both directions (increase and decrease) opens up new possibilities for frequency detuning of superconducting qubits. The developed method of thermal annealing allows not only to increase the junction resistance by 175%, but also to decrease $R_n$ by 60%. The tuning technique is capable of precisions of 10% in $R_n$ as the resistance decreases. LASIQ allows you to point-to-point increase in resistance, lowering the qubits in frequency. While our method allows to roughly reduce the resistance by raising the frequency value, and then locally tune them.

## Conclusions

In an effort to solve the problems of frequency collisions and decoherence of super-conducting qubits, we have performed a systematic study of the technology of thermal annealing of Josephson junctions. To do this, we fabricated a significant number of junctions and directly measured the change in their room temperature resistance depending on the thermal annealing mode. We achieved Al/AlO$_x$/Al junction $R_n$ increase by 175% and decrease by 60% with precisions of 10% in $R_n$. Theoretical assumptions about modification of the barrier structure are proposed. The results obtained open up new possibilities for frequency tuning of qubits: raise or lower the frequencies of already fabricated qubits across the chip. Proposed method is compatible with LASIQ technology, and their combination will allow tuning the qubit frequencies over a wide range.

## Data availability

The datasets used and/or analysed during the current study available from the corresponding author on reasonable request.

## Acknowledgements

Technology was developed and samples were fabricated at the BMSTU Nanofabrication Facility (Functional Micro/Nanosystems, FMNS REC, ID 74300).

## Author contributions statement

I.R. (Ilya Rodionov), I.A.R. (Ilya Anatolevich Ryzhikov), D.O.M. conceptualized the ideas of the project. N.D.K., A.A.S. and M.V.A. fabricated experimental samples and discussed results. D.A.M. and A.R.I. performed morphology characterization. N.D.K. and E.S.L. conducted the process of thermal annealing of the experimental samples. N.D.K. and D.O.M. conducted the electrical characterization of the experimental samples. N.D.K., D.O.M., I.A.R. and I.R. analyzed the experimental data and discussed the results. N.D.K., D.O.M. and Y.V.P. prepared writing-original draft. I.R. reviewed and edited the manuscript. I.R. supervised the project. All authors analyzed the data and contributed to writing the manuscript.

## Additional information

The authors declare no conflict of interest.